\documentclass[twocolumn]{aastex631}

\usepackage{multirow}

\shorttitle{Average [CII] $158\,\rm{\mu m}$ sizes of Star-Forming Galaxies between $z\sim7$ and $z\sim4$}
\shortauthors{Fudamoto et al.}

\begin{document}

\title{The ALMA REBELS Survey: Average  [CII] $\mathbf{158\,\mathbf{\mu m}}$ sizes of Star-Forming Galaxies from $\mathbf{z\sim7}$ to $\mathbf{z\sim4}$}

\author[0000-0001-7440-8832]{Y. Fudamoto} 
\affiliation{Waseda Research Institute for Science and Engineering, Faculty of Science and Engineering, Waseda University, 3-4-1 Okubo, Shinjuku, Tokyo 169-8555, Japan}
\affiliation{National Astronomical Observatory of Japan, 2-21-1, Osawa, Mitaka, Tokyo, Japan}

\author[0000-0001-8034-7802]{R. Smit} 
\affiliation{Astrophysics Research Institute, Liverpool John Moores University, 146 Brownlow Hill, Liverpool L3 5RF, United Kingdom}

\author[0000-0002-7252-4589]{R. A. A. Bowler} 
\affiliation{Jodrell Bank Centre for Astrophysics, Department of Physics and Astronomy, School of Natural Sciences, The University of Manchester, Manchester, M13 9PL, UK}

\author[0000-0001-5851-6649]{P. A. Oesch} 
\affiliation{Observatoire de Gen{\`e}ve, 1290 Versoix, Switzerland}
\affiliation{Cosmic Dawn Center (DAWN), Niels Bohr Institute, University of Copenhagen, Jagtvej 128, K{\o}benhavn N, DK-2200, Denmark}

\author{R. Bouwens} 
\affiliation{Leiden Observatory, Leiden University, NL-2300 RA Leiden, Netherlands}

\author{M. Stefanon} 
\affiliation{Leiden Observatory, Leiden University, NL-2300 RA Leiden, Netherlands}

\author[0000-0003-4268-0393]{H. Inami} 
\affiliation{Hiroshima Astrophysical Science Center, Hiroshima University, 1-3-1 Kagamiyama, Higashi-Hiroshima, Hiroshima 739-8526, Japan}

\author[0000-0003-4564-2771]{R. Endsley} 
\affiliation{Steward Observatory, University of Arizona, 933 N Cherry
Ave, Tucson, AZ 85721, United States}

\author[0000-0002-3120-0510]{V. Gonzalez} 
\affiliation{Departmento de Astronomia, Universidad de Chile, Casilla
36-D, Santiago 7591245, Chile}
\affiliation{Centro de Astrofisica y Tecnologias Afines (CATA), Camino
del Observatorio 1515, Las Condes, Santiago, 7591245, Chile}

\author{S. Schouws} 
\affiliation{Leiden Observatory, Leiden University, NL-2300 RA Leiden, Netherlands}

\author{ D. Stark} 
\affiliation{Steward Observatory, University of Arizona, 933 N Cherry
Ave, Tucson, AZ 85721, United States}

\author[0000-0002-4205-9567]{H. S. B. Algera} 
\affiliation{Hiroshima Astrophysical Science Center, Hiroshima University, 1-3-1 Kagamiyama, Higashi-Hiroshima, Hiroshima 739-8526, Japan}
\affiliation{National Astronomical Observatory of Japan, 2-21-1, Osawa, Mitaka, Tokyo, Japan}

\author[0000-0002-6290-3198]{M. Aravena} 
\affiliation{Nucleo de Astronomia, Facultad de Ingenieria y Ciencias,
Universidad Diego Portales, Av. Ejercito 441, Santiago, Chile}

\author{L. Barrufet} 
\affiliation{Observatoire de Gen{\`e}ve, 1290 Versoix, Switzerland}

\author{E. da Cunha} 
\affiliation{International Centre for Radio Astronomy Research, University of Western Australia, 35 Stirling Hwy, Crawley,26WA
6009, Australia}

\author{P. Dayal} 
\affiliation{Kapteyn Astronomical Institute, University of Groningen, P.O. Box 800, 9700 AV Groningen, The Netherlands}

\author{A. Ferrara} 
\affiliation{Scuola Normale Superiore, Piazza dei Cavalieri 7, 56126
Pisa, Italy}

\author[0000-0002-9231-1505]{L. Graziani} 
\affiliation{Dipartimento di Fisica, Sapienza, Universita di Roma, Piazzale Aldo Moro 5, I-00185 Roma, Italy}
\affiliation{INAF/Osservatorio Astrofisico di Arcetri, Largo E. Femi 5,
I-50125 Firenze, Italy}

\author{J. A. Hodge} 
\affiliation{Leiden Observatory, Leiden University, NL-2300 RA Leiden, Netherlands}

\author[0000-0002-6488-471X]{A. P. S. Hygate} 
\affiliation{Leiden Observatory, Leiden University, NL-2300 RA Leiden, Netherlands}

\author[0000-0002-7779-8677]{A. K. Inoue} 
\affiliation{Waseda Research Institute for Science and Engineering, Faculty of Science and Engineering, Waseda University, 3-4-1 Okubo, Shinjuku, Tokyo 169-8555, Japan}
\affiliation{Department of Physics, School of Advanced Science and Engineering, Faculty of Science and Engineering, Waseda University, 3-4-1, Okubo, Shinjuku, Tokyo 169-8555, Japan}

\author{T. Nanayakkara} 
\affiliation{Centre for Astrophysics \& Supercomputing, Swinburne University of Technology, PO Box 218, Hawthorn, VIC 3112, Australia}

\author[0000-0002-7129-5761]{A. Pallottini} 
\affiliation{Scuola Normale Superiore, Piazza dei Cavalieri 7, 56126
Pisa, Italy}

\author[0000-0002-9712-0038]{E. Pizzati} 
\affiliation{Scuola Normale Superiore, Piazza dei Cavalieri 7, 56126
Pisa, Italy}
\affiliation{Leiden Observatory, Leiden University, NL-2300 RA Leiden, Netherlands}

\author[0000-0001-9317-2888]{R. Schneider} 
\affiliation{Dipartimento di Fisica, Sapienza, Universita di Roma, Piazzale Aldo Moro 5, I-00185 Roma, Italy}
\affiliation{INAF/Osservatorio Astronomico di Roma, via Frascati 33,
00078 Monte Porzio Catone, Roma, Italy}

\author{L. Sommovigo} 
\affiliation{Scuola Normale Superiore, Piazza dei Cavalieri 7, 56126
Pisa, Italy}

\author[0000-0001-6958-7856]{Y. Sugahara} 
\affiliation{Waseda Research Institute for Science and Engineering, Faculty of Science and Engineering, Waseda University, 3-4-1 Okubo, Shinjuku, Tokyo 169-8555, Japan}
\affiliation{National Astronomical Observatory of Japan, 2-21-1, Osawa, Mitaka, Tokyo, Japan}

\author{M. Topping} 
\affiliation{Steward Observatory, University of Arizona, 933 N Cherry
Ave, Tucson, AZ 85721, United States}

\author[0000-0001-5434-5942]{P. van der Werf} 
\affiliation{Leiden Observatory, Leiden University, NL-2300 RA Leiden, Netherlands}

\author[0000-0002-3915-2015]{M. Bethermin} 
\affiliation{Aix Marseille Universit{\'e}, CNRS, LAM (Laboratoire d’Astrophysique de Marseille), F-13388 Marseille, France}
\affiliation{Universite de Strasbourg, CNRS, Observatoire astronomique de Strasbourg, UMR 7550, 67000 Strasbourg, France}

\author{P. Cassata} 
\affiliation{Dipartimento di Fisica e Astronomia Galileo Galilei Universit{\`a} degli Studi di Padova, Vicolo dell’Osservatorio 3, 35122 Padova Italy}

\author{M. Dessauges-Zavadsky} 
\affiliation{Observatoire de Gen{\`e}ve, 1290 Versoix, Switzerland}

\author[0000-0002-9382-9832]{A. L. Faisst} 
\affiliation{IPAC, California Institute of Technology, 1200 East California Boulevard, Pasadena, CA 91125, USA}

\author[0000-0001-7201-5066]{S. Fujimoto} 
\affiliation{Cosmic Dawn Center (DAWN), Jagtvej 128, DK2200, Copenhagen N, Denmark}
\affiliation{Niels Bohr Institute, University of Copenhagen, Lyngbyvej 2, DK2100 Copenhagen, Denmark}

\author[0000-0002-9122-1700]{M. Ginolfi} 
\affiliation{European Southern Observatory, Karl-Schwarzschild-Str. 2, D-85748 Garching bei M{\"u}nchen, Germany}

\author[0000-0001-6145-5090]{N. Hathi} 
\affiliation{Space Telescope Science Institute, 3700 San Martin Drive, Baltimore, MD 21218, USA}

\author{G. C. Jones} 
\affiliation{Department of Physics, University of Oxford, Denys Wilkinson Building, Keble Road, Oxford OX1 3RH, UK}

\author{F. Pozzi} 
\affiliation{Dipartimento di Fisica e Astronomia, Universit{\`a} of Bologna, via Gobetti 93/2, 40129, Bologna, Italy}
\affiliation{INAF/Osservatorio di Astrofisica e Scienza dello Spazio di Bologna, via Gobetti 93/3, 40129, Bologna, Italy}

\author[0000-0001-7144-7182]{D. Schaerer} 
\affiliation{Observatoire de Gen{\`e}ve, 1290 Versoix, Switzerland}



\begin{abstract}
We present the average [CII] $158\,\rm{\mu m}$ emission line sizes of UV-bright star-forming galaxies at $z\sim7$.
Our results are derived from a stacking analysis of [CII] $158\,\rm{\mu m}$ emission lines and dust continua observed by ALMA, taking advantage of the large program Reionization Era Bright Emission Line Survey (REBELS).
We find that the average [CII] emission at $z\sim7$ has an effective radius $r_e$ of $2.2\pm0.2\,\rm{kpc}$. It is $\gtrsim2\times$ larger than the dust continuum and the rest-frame UV emission, in agreement with recently reported measurements for $z\lesssim6$ galaxies.
Additionally, we compared the average [CII] size with $4<z<6$ galaxies observed by the ALMA Large Program to INvestigate [CII] at Early times (ALPINE).
By analysing [CII] sizes of $4<z<6$ galaxies in two redshift bins, we find an average [CII] size of $r_{\rm e}=2.2\pm0.2\,\rm{kpc}$ and $r_{\rm e}=2.5\pm0.2\,\rm{kpc}$ for $z\sim5.5$ and $z\sim4.5$ galaxies, respectively.
These measurements show that star-forming galaxies, on average, show no evolution in the size of the [CII] $158\,{\rm \mu m}$ emitting regions at redshift between $z\sim7$ and $z\sim4$.
This finding suggest that the star-forming galaxies could be morphologically dominated by gas over a wide redshift range.
\end{abstract}



\section{Introduction}
Investigating star formation activity in the early Universe is key to understand galaxy formation and evolution.
Thanks to deep galaxy surveys with the Hubble Space Telescope (HST) and large ground-based telescopes, it is now widely established that high-redshift galaxies (from $z\sim11$ to $z\sim4$) have been rapidly forming stars at an accelerating rate \cite[e.g.,][]{Madau2014}, supported by high gas fractions (e.g., \citealt{Dayal2014,Liu2019,Decarli2020,Dessauges2020,Wang2022}, and see \citealt{Tacconi2020}, for a review).
Investigating the gas supply that fuels the star formation activity requires detailed studies of the spatial distribution of gas within and/or around galaxies.
However, this is still poorly understood as detailed observations of the interstellar medium or circumgalactic medium at high redshift have been limited.

In recent years, the Atacama Large Millimeter/submillimeter Array (ALMA) made it possible to observe ISM properties of  high-redshift galaxies in great detail.
In particular, with its unprecedented sensitivity, ALMA provided us with extremely deep surveys of high-redshift galaxies \citep[see][for a review]{Hodge2020}.
These ALMA observations revealed that the spatial distributions of interstellar gas seen through the far-infrared (FIR) emission line [CII] $158\,{\rm \mu m}$ is more extended than the dust continuum and rest-UV emission \citep[e.g.,][]{Fujimoto2019,Fujimoto2020,Herrera-Camus2021}.
Previous studies have suggested that these extended gas reservoirs are ubiquitous in high-redshift ($z>5$) star-forming galaxies, and are linked to outflow features \citep[e.g.,][]{Gallerani2018,Ginolfi2020,Graziani2020,Pizzati2020}.
However, these features are not yet confirmed for $z>6$ star-forming galaxies, and it is not clear if the extended gas properties systematically change as a function of redshift.

In this paper, we investigate the average size of the [CII] $158\,\rm{\mu m}$ emission line and dust continua of $z\sim7$ galaxies based on the on-going ALMA large program Reionization Era Bright Emission Line Survey \citep[REBELS;][]{Bouwens2021}. We compare the $z\sim7$ size measurements with observations of $z\sim4-6$ galaxies from the ALMA Large Program to INvestigate [CII] at Early times \citep[ALPINE;][]{Lefevre2020,Bethermin2020,Faisst2020} to investigate if the spatial distribution of the ISM between these two redshift ranges.

This paper is organized as follows: in \S2 we describe our observations and the sample used in this study. In \S3, we present our methodology for stacking and size measurements.  \S4 shows the results and discussion on the stacked [CII] emission and dust continuum.
Throughout this paper, we assume a cosmology with $(\Omega_m,\Omega_{\Lambda},h)=(0.3,0.7,0.7)$, and the Chabrier \citep{Chabrier2003} initial mass function (IMF), where applicable.
With these cosmological parameters, 1 arcsec corresponds to $6.28\,\rm{pkpc}$ and $5.23\,\rm{pkpc}$ at $z=5$ and $z=7$, respectively.

\section{Data} \label{sec:data}

\subsection{Sample and ALMA Observations} \label{sec:ALMAobs}
Our analysis of $z\sim7$ galaxies is based on observations of the [CII] $158\,\rm{\mu m}$ line from the ALMA large program REBELS (PID: 2019.1.01634.L).
REBELS used spectral line scans to search for the [CII] $158\,\rm{\mu m}$ line in 36 galaxies and the [OIII] $88\,\rm{\mu m}$ line in 4 galaxies.
In this study, we use the 34 completed observations from Cycle-7, targeting [CII] emission lines, in UV-selected galaxies from $z=6.5$ to $z=9$.
These scans were carried out in band-5 or band-6 using compact configurations (C43-1 and C43-2), resulting in the typical synthesized beam full width at half maximum (FWHM) of $\sim1.2-1.6^{\prime\prime}$.
We refer to \citet{Bouwens2021}, Schouws et al. (in prep.), and \citet{Inami2022} for a complete description of the survey, ALMA data processing, and dust continuum detections, respectively.
In addition to REBELS, we include 8 additional $z>6.5$ galaxies from pilot ALMA [CII] observations (PID: 2015.1.01111.S, 2018.1.00085.S, 2018.1.00236.S).
These additional observations employ identical sample selection criteria, spectral scan strategy, and angular resolution to the REBELS survey \citep[see ][for details]{Smit2018,Schouws2021,Schouws2022}.
In total, we consider 42 separate ALMA targets as part of this analysis (2 sources are in common between REBELS and the pilot programs).

These observations, in summary, target UV-bright star-forming galaxies at $z\sim7$.
The target galaxies consist of the brightest ($-23\leq M_{\rm UV}<-21.4$) and highest mass ($8.6 < {\rm log\,}(M_{\ast}/{\rm M_{\odot}}) < 10.1$; Stefanon et al. in prep.) star-forming galaxies at $z\sim7$ \citep{Bouwens2021}.

Additionally, we complemented our sample with the ALMA survey targeting $z\sim4.5$ to $z\sim6$ galaxies \citep[ALPINE survey: ][]{Lefevre2020}, as a lower redshift comparison sample. 
The ALPINE survey targeted 118 UV-bright main-sequence galaxies, spanning a stellar mass range $8.4 < {\rm log\,}(M_{\ast}/{\rm M_{\odot}}) < 11.0$ and UV magnitudes of $-23.3<M_{\rm UV}<-19.2$ \citep{Faisst2020}.
These galaxies are the ideal comparison sample at $4<z<6$, as the ALPINE survey provides the largest and the most homogeneous data set of [CII] $158\,\rm{\mu m}$ emission lines and dust continua of $4<z<6$ star-forming galaxies.
We refer to \citet{Lefevre2020}, \citet{Bethermin2020}, and \citet{Faisst2020} for a complete description of the survey objectives, the ALMA data processing, and the multiwavelength ancillary observations, respectively.
The data are available publicly on the ALPINE website\footnote{\url{https://cesam.lam.fr/a2c2s/}}.

\subsection{ALMA Detections}

For our [CII] emission stacking analysis at $z\sim7$, we use 28 individually detected [CII] emission lines (signal to noise ratio; SNR $\gtrsim 5.2$) from REBELS survey (23 galaxies) and pilot observations (5 galaxies). For the continuum stacking analysis at $z\sim7$, we include 16 individual dust continuum detections (at SNR $> 3.3$) from REBELS survey (14 galaxies) and pilot observations (2 galaxies). The detection threshold is sufficient to guarantee a $\geq 95\%$ purity for the [CII] emission lines and dust continua (\citealt{Inami2022}; Schouws et al. in prep). The [CII] emission and dust continuum stacks are made based on individual detections of each emission. Specifically, we did not include galaxies that have [CII] detection but no continuum detection when constructing the continuum stack. This stacking strategy allows us to produce the highest SNR stacks as well as to avoid additional uncertainty of continuum size measurements, potentially arising from unknown continuum position (see \S\ref{sec:stack} as well). 

The detected [CII] lines have luminosities in the range between $8.1 < {\rm log\,}(L_{\rm [CII]}/{\rm L_{\odot}}) < 9.2$ with a median of ${\rm log\,}(L_{\rm [CII]}/{\rm L_{\odot}}) = 8.8$ (Schouws et al. in prep.).
The continuum luminosities are estimated using a median conversion factor of $L_{\rm{IR}}=14^{+8}_{-5}\,\nu\,L_{\nu,158\,\rm{\mu m}}$  based on the infrared SED derived by \citet{Sommovigo2022}. The estimated dust continuum luminosities have a range of $11.5<{\rm log} (L_{\rm IR}/{\rm L_{\odot}}) <12.2$ with a median of ${\rm log} (L_{\rm IR}/{\rm L_{\odot}}) = 11.6$ \citep{Inami2022}.
While the average offset between detected [CII] emission and dust continuum is $\sim0.35^{\prime\prime}$ for our $z\sim7$ galaxies (i.e., well within the synthesized beam size), two galaxies (REBELS-12 and REBELS-19) show much larger spatial offsets ($\gtrsim1^{\prime\prime}$).
As these galaxies are  potentially on-going mergers \citep{Inami2022}, we removed these two  from our analysis, leaving 26 [CII] lines and 14 dust continuum detections.

For the $4<z<6$  galaxies, we selected galaxies from the ALPINE public catalog \citep{Bethermin2020}, and included galaxies individually detected in [CII] for our analysis.
The [CII] detection threshold of ALPINE (SNR$>3.5$) corresponds to 95\% purity, similar to our $z\sim7$ galaxies, ensuring that both observations have little spurious source contamination.
Furthermore, to avoid on-going galaxy mergers contaminating our morphology analysis, we excluded [CII] emission lines showing merger events based on the morpho-kinematic classification by \citet{Romano2021}.
These selection provides 52 galaxies from ALPINE survey: 31 for $z\sim4.5$ and 21 for $z\sim5.5$.

\section{Analysis} \label{sec:analysis}

\subsection{Stacking ALMA Images} \label{sec:stack}

To investigate the average [CII] and dust continuum sizes, we made stacked images of the [CII] $158\,\rm{\mu m}$ emission lines, and also dust continua of $z\sim7$ galaxies.
In the stacks, we included only individually detected [CII] emission lines and dust continua.

We note that stacking non-detected [CII] emission is difficult as [CII] non-detected galaxies only have photometric redshift.
For dust continua, it would be still possible to include individually non-detected continuum.
Nevertheless, we stacked only individually detected continua to make the highest SNR images, enabling a detailed study of the average [CII] and dust morphology.
At the same time, this method helps to avoid possible systemic uncertainty of stacked sizes of dust continuum arising from unknown positions of the individually non-detected emissions. In particular, peak positions of detected [CII] and dust continua show $\sim0.35^{\prime\prime}$ of offsets on average (\citealt{Inami2022}; Schouws et al., in prep). Thus, stacking non-detected continuum could introduce such systemic uncertainty on the measured small size (see \ref{sec:sizehighz}).

For the stacks, we start with the $35^{\prime\prime}\times35^{\prime\prime}$ moment-0 maps that were made by integrating over the $2\,\sigma$ velocity width of the [CII] emission lines after continuum subtraction, while the continuum images were made by removing channels that are in the $3\,\sigma$ velocity widths of the detected [CII] emission lines.
While the $2\,\sigma$ velocity integration of [CII] could miss some of the high velocity, faint component arising from outflowing gas \citep[e.g.,][]{Ginolfi2020}, we decided the integration velocity width to focus on galaxy's most [CII]-bright component (i.e., host galaxies of outflows if they exist). This is in-line with the previous study that studied  ``core'' component of [CII] emission by selecting $\pm50\,\rm{km/s}$ velocity width of [CII] \citep{Fujimoto2019}.
To avoid artifacts, these moment-0 maps were not deconvolved with the synthesized beam (i.e., without cleaning).
After centering images to each of the peak fluxes, these maps were average-stacked using an inverse variance weighting, where the variance is measured using the background RMS of each map. We derived effective synthesized beams of the stacks by weighted-averaging all the dirty beams employing the same weights as for the moment-0 images.

We examined if the image-based stacking method systematically affects our results by comparing with the visibility-based stack.
We performed this test using $z\sim7$ galaxies.
We stacked the [CII] visibility data following the methods of \citet{Fujimoto2019}. We then measured stacked [CII] sizes using the visibility-based fitting software \path{UVMULTIFIT} \citep{Marti-Vidal2014}.
The resulting fits are shown in lower panels of Fig. \ref{fig:galfit}.
We found both stacked images are almost identical, and size measurements from both images agree well within $<7\,\%$. Given that making stacked visibility data, especially the data concatenations (i.e., \path{concat} task in CASA), is time-expensive, and given both methods provide consistent results, we use the image based stacking in the following analysis. In our case, using image based stacking helps to produce a lager number of stacked images required in the bootstrap.

To check if a small fraction of extreme galaxies in our sample could bias our measurements, we performed a bootstrap analysis to estimate the uncertainties coming from both the noise and the sample variance.
We made 1000 stacks using randomly selected $N$ galaxies allowing overlaps, where $N$ is the number of galaxies in the original stack.
Throughout this paper the reported measurements are the median of the bootstrap resampling, and uncertainties are based on the 16th and the 84th percentiles.

\subsection{Stacking the Rest-Frame UV Images} \label{sec:restUVstack}
To examine sizes of $z\sim7$ galaxies in different wavelengths, we performed a stacking analysis of the rest-frame UV images using the method of \citet{Bowler2017}.
Although high resolution observations, such as using HST, is required to provide secure constraints, only a small subset of our sources (4 galaxies) have HST observations \citep{Bowler2022}. 
To provide tentative limits of rest-frame UV sizes, we used ground based observations; publicly available J-band images from the Ultra VISTA survey \citep{McCracken2012} and the VIDEO survey \citep{Jarvis2013}; the resulting stack has a point spread function FWHM of $\sim0.9^{\prime\prime}$. Same as [CII] and continuum stack, we performed a bootstrap analysis to estimate the certainty of the rest-frame UV sizes.  A possible caveat of only using ground base observations is discussed in \S\ref{sec:caveats}.

For $z<6$ galaxies, we used rest-UV size measurements from \citet{Fujimoto2020}, which uses deep HST F160W images.

\subsection{Size Measurements}
\begin{figure}[tb!]
\centering
\includegraphics[width=\columnwidth]{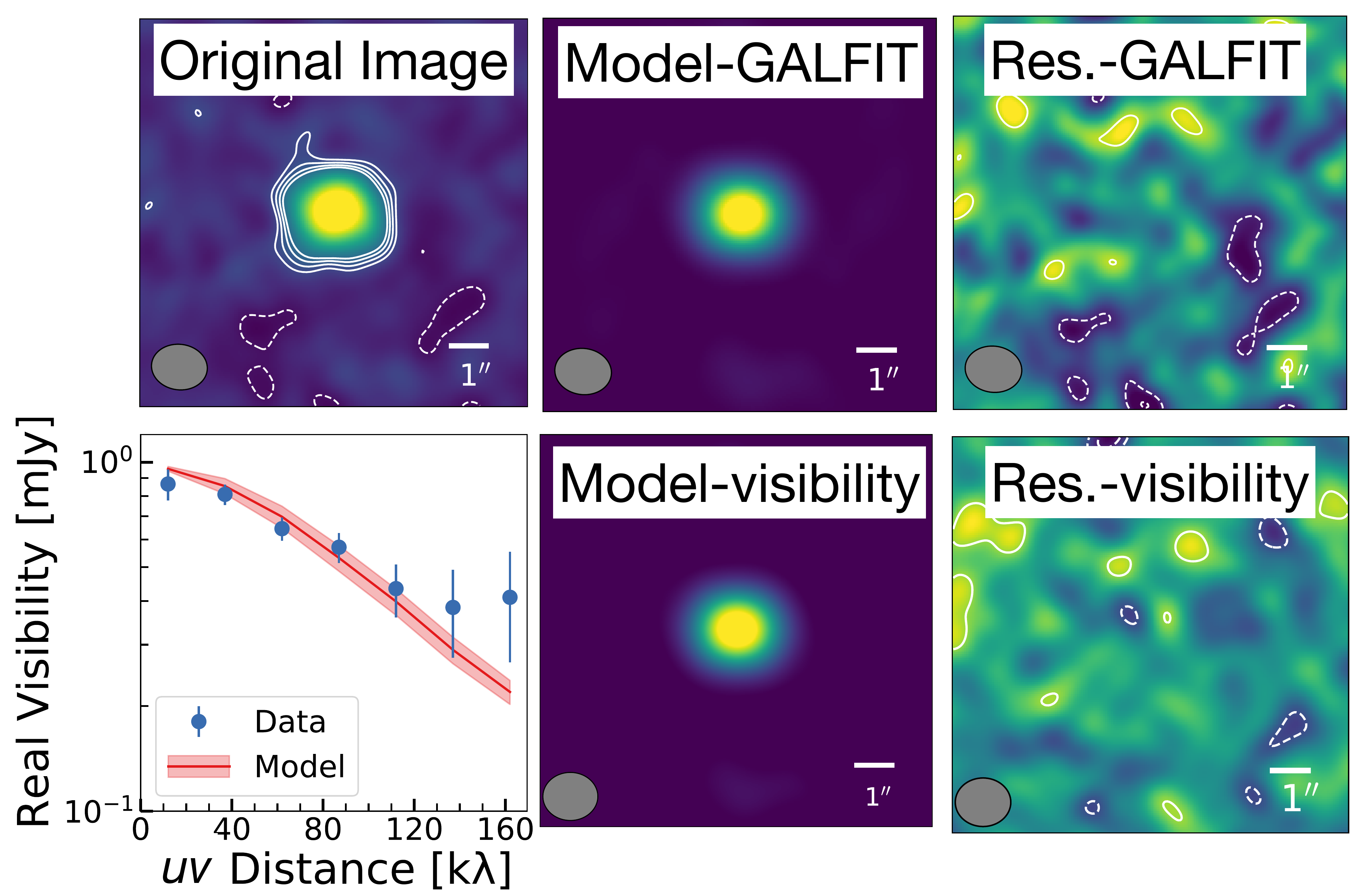}
\caption{[{\bf Upper Panels}]: An example of the size measurements we make of the [CII] emission using GALFIT \citep{Pang2010}. The upper left, middle, and right panels show a stacked image of [CII] emission resulting from our bootstrap resampling process, the beam convolved model image, and the residual image, respectively.
[{\bf Lower Panels}]: An example of using UVMODELFIT \citep{Marti-Vidal2014} to the visibilities of the stacked [CII] emission. The lower left panel shows the best fit visibility model (red line with $1\,\sigma$ band). Blue points are the median of the $25\,\rm{k\lambda}$ averaged data. Error bars shows standard deviation of the data. The lower middle and right panels show model and residual of the results of fitting in the visibility data. These figures show that the stacked [CII] emission is well resolved, and our measurements provide robust results.
White solid (dashed) contours show 2,3,4,$5\,\sigma$ ($-2\,\sigma$) signal, and ellipses in the lower left corner show the stacked synthesized beam. 
All residual images show no large negative or positive signal, showing that our fits are successful.
\label{fig:galfit}
}
\end{figure}

We used GALFIT \citep{Pang2010} to measure the beam-deconvolved effective radius ($r_{\rm e}$) on the stacked maps of the [CII], dust continuum, and the rest-frame UV emission.
For each GALFIT run, we assumed an exponential disk surface brightness profile, similar to previous studies \citep[e.g.,][]{Fujimoto2019,Fujimoto2020}, which assume that the [CII] surface brightness profile traces the gas distribution of galaxies \citep[e.g.,][]{Bigiel2012}. The exponential disk profile is in the form of $\propto{\rm exp}(-r/r_{\rm s})$ where $r_{\rm s}$ is the scale length of the exponential profile, and $r_{\rm s}$ can be converted to the effective radius $r_{\rm e}$ by $r_{\rm e}=1.678\,r_{\rm s}$ \citep{Pang2010}.
We also fixed the axial ratio to be 1 (i.e., circular exponential profiles) as stacks are expected to average over randomly oriented galaxies.
Using the stacked synthesized images, we find that the [CII] emission line sizes are well resolved and constrained.
Fig. \ref{fig:galfit} shows an example of our fitting results.

\section{Results and Discussion}

\begin{figure}[tb!]
\centering
\includegraphics[width=8cm]{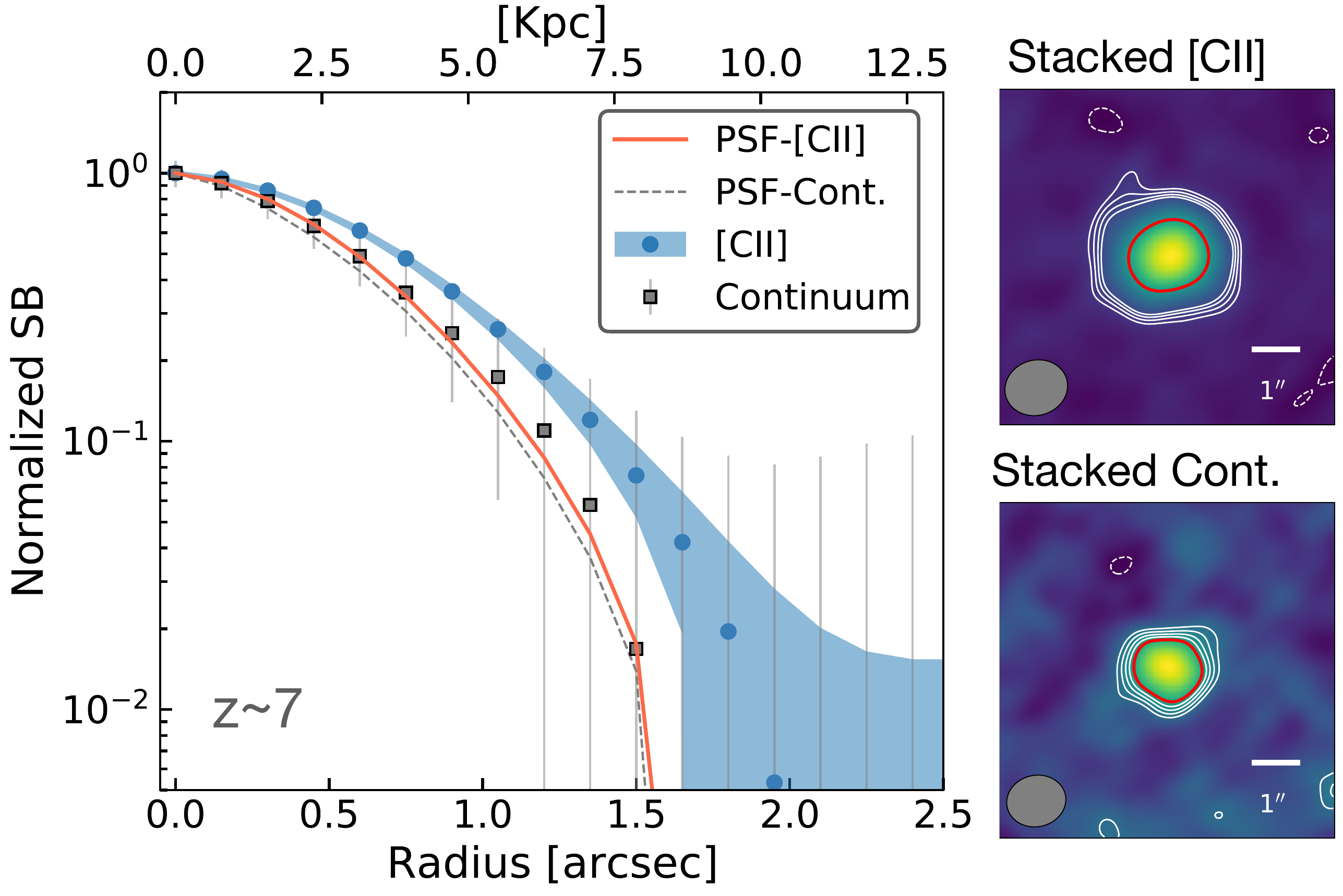}
\caption{
    {\it Left panel:} Normalized surface brightness profiles of the stacked [CII] emission at $z\sim7$ (blue points with shaded region), stacked dust continuum  (gray squares with error bars), and synthesized beams at $z\sim7$ (for [CII] in the orange line; for dust continuum for dashed gray line).
    The normalized surface brightness values are median values within annuli having $0.125^{\prime\prime}$ widths centred on the peak of emissions.
    The normalized standard deviations of the [CII] and continuum profiles are shwon by the shaded area and error bars, respectively. These surface brightness become less than $1\,\sigma$ at $r>1^{\prime\prime}$ and at $1.6^{\prime\prime}$ for continuum and [CII] emission, respectively.
    While the stacked continuum is only barely resolved, the stacked [CII] emission is well resolved and has a physical extent of $r_{\rm e}\sim 2.2\,\rm{kpc}$. This shows that the [CII] emitting region is significantly more extended than the continuum region.
    {\it Right upper panel:} Stacked [CII] $158\,\rm{\mu m}$ emission line map at $z\sim7$.
    {\it Right lower panel:} Stacked dust continuum map map at $z\sim7$.
    In the maps, contours show $3,4,5,6\,\sigma$ (solid) and $-3\,\sigma$ (dashed).
    Red contours show the half maximum power of the emission.
    Synthesized beams are shown in the lower left corners.
    \label{fig:profile}
    }
\end{figure}

\subsection{Dust and [CII] Sizes of $z\sim7$ Galaxies}
\label{sec:sizehighz}

We studied radial profiles of the stacked dust continuum and [CII] emission. We first measured surface brightness profiles of the emissions by calculating the median value within annuli of width $0.125^{\prime\prime}$ centred on the emission peaks. Errors are estimated using background standard deviation of the stacked images. We then measured surface brightness of the stacked synthesized beams using same method to check how well the stacked emission are resolved. By comparing normalized surface brightness, we compare relative extensions of continuum and [CII] emission (Fig. \ref{fig:profile}).

The radial profile of the stacked dust continuum is consistent with the stacked synthesized beam. This indicates that the dust continuum is, on-average, much smaller than the current spatial resolution of $\sim1.3^{\prime\prime}$.
Using GALFIT, we find the dust continuum effective radius of $r_{\rm{e-cont.}}=1.14\pm0.27\,\rm{kpc}$ ($0.22\pm0.06\,\rm{arcsec}$),  where $r_{\rm{ e-cont.}}$ is estimated by the median GALFIT outputs.
We note, however, that the estimated $r_{\rm{ e-cont.}}$ is highly uncertain as continua are only barely resolved using the synthesized beam FWHM of $\sim1.3^{\prime\prime}$. We treat the estimated continuum size as a tentative measurement in the following, and we focus only on [CII] size comparisons with $z\sim4-6$ galaxies (\S 4.2).
Higher resolution observations are required to measure continuum sizes more accurately.

The stacked [CII] emission line is, on the other hand, well resolved and is clearly more extended than the synthesized beam (Fig. \ref{fig:profile}).
The stacked profile can be fitted with a single [CII] exponential component, without the need for the second, more concentrated one introduced in  \citet{Fujimoto2019}. Such a difference might result from the lower angular resolution ($\sim1.3^{\prime\prime}$) of our observations, which corresponds to a typical beam size $\gtrsim2\times$  larger than the \citet{Fujimoto2019} one.
As a result of the bootstrap resampling, we find the [CII] effective radius of $r_{\rm e}=2.21^{+0.23}_{-0.19}\,\rm{kpc}$ ($0.42\pm0.04\,\rm{arcsec}$) for $z\sim7$ galaxies, where $r_{\rm e}$ and errors are estimated in the same manner as the continuum size.

To study rest-frame UV sizes of $z\sim7$ galaxies, we also used GALFIT to fit to the exponential disk profile to the stacked image (see \S\ref{sec:restUVstack}). We found the stacked rest-frame UV image has an effective radius of   $r_{\rm{e-UV}}=0.83\pm0.16\,\rm{kpc}$.
The result shows that the rest-frame UV sizes of our sample are, on average, $\sim3\times$ smaller than the [CII] sizes.
We note, however, that the stacked rest-frame UV image is only marginally resolved as only ground-based data are available (see  \citealt{Bowler2022} for HST observations of a subset of the REBELS targets).
We report the rest-frame UV size as a tentative measurement.
See \S\ref{sec:caveats} for discussions about possible systematic uncertainties of our rest-frame UV size measurement.

The extended [CII] emission and the compact dust continuum show that at $z\sim7$ the [CII] emitting gas is, on average, more extended than the star-forming regions.
These results are consistent with previous findings using star-forming galaxies at $z\sim4-6$ \citep{Fujimoto2019,Fujimoto2020,Herrera-Camus2021}.
Combined with previous studies, our results show that the [CII] emitting gas is spatially more extended than actively star-forming regions over a wide redshift range.

\subsection{Non-Evolution of [CII] Sizes from $z\sim7$ to $z\sim4.5$}
\label{sec:nonevolution}

\begin{figure}[bt!]
\centering
\includegraphics[width=8cm]{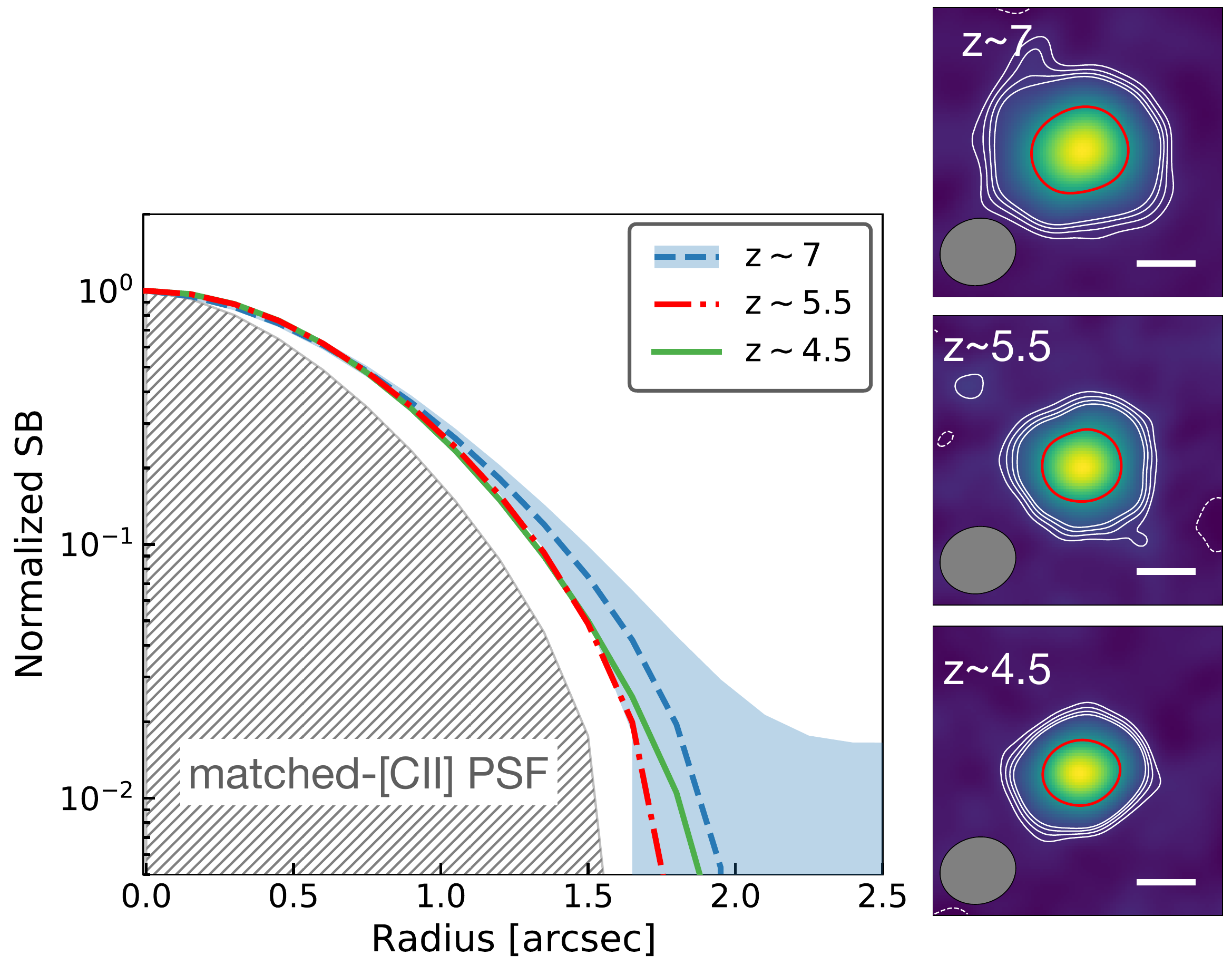}
\caption{
    {\it Left Panel:} The stacked radial profiles of the beam matched [CII] $158\,\rm{\mu m}$ emission from galaxies in three redshift bins. The lines show median profiles of galaxies at $z\sim7$ (blue dashed), $z\sim5.5$ (red dashed dot), and $z\sim4.5$ (green solid), and the hatched profile shows the matched beam.
    The normalized surface brightness is measured using annuli having $0.125^{\prime\prime}$ widths centred at the peak of emissions. The band show normalized standard deviations of the background images.
    All stacked profiles have significant excesses from the stacked beams, showing that stacked [CII] lines are well resolved spatially, and all profiles have consistent angular sizes and shapes.
    {\it Right Panels:} $5^{\prime\prime}\times5^{\prime\prime}$ stamps of the stacked [CII] moment-0 maps in different redshift bins. White solid (dotted) contours show 2 to 5 $\,\sigma$ (-2$\,\sigma$) signals if present.
    White bars in the lower right corners show $1^{\prime\prime}$ scales.
    Red contours show the half maximum power of the emission.
    \label{fig:profile_compare}
    }
\end{figure}

\begin{figure}[bt!]
\centering
\includegraphics[width=8cm]{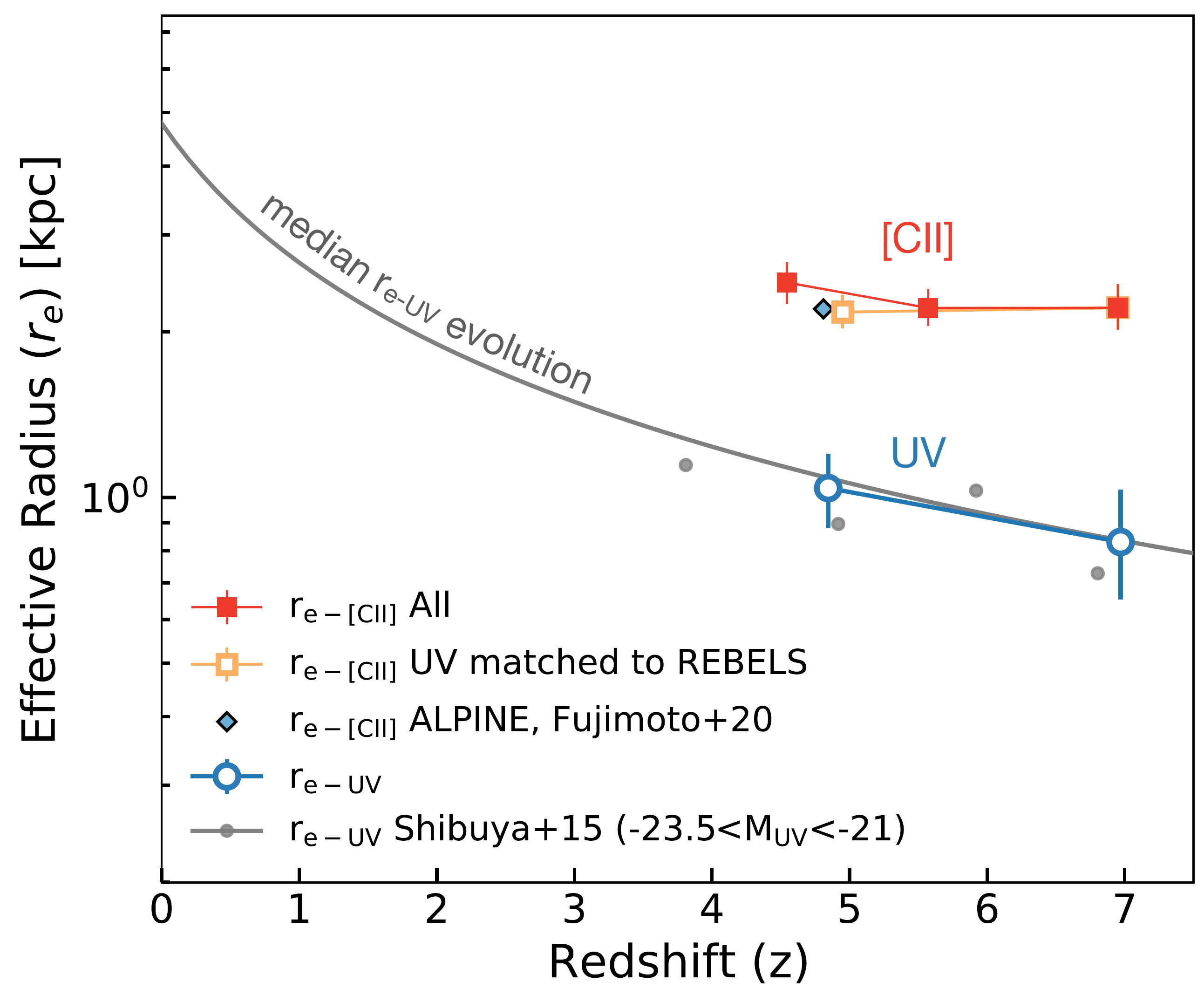}
\caption{
    Effective radius ($r_e$) as a function of redshift as measured from the [CII] $158\,\rm{\mu m}$ emission line and rest-UV continuum.
    The effective radius of [CII] emission (red squares) shows little change between $z\sim4.5$ and $z\sim7$, while the rest-UV continuum, on average, is smaller at higher redshift (gray dots and line from \citealt{Shibuya2015}, and blue open circles from \citealt{Fujimoto2020} and this work). The non-evolution of the [CII] size at $z\sim4-7$ remains present when we use a UV luminosity matched sample of galaxies at $z\sim5$ (yellow open square). This suggests that high-redshift galaxies might be morphologically dominated by gas, and that star formation activity occupies a progressively smaller fraction of the volume in galaxies towards the highest redshifts.
    \label{fig:size-evolution}
    }
\end{figure}

Similar to $z\sim7$ galaxies, the stacked [CII] emission of $z\sim4.5$ and $z\sim5.5$ galaxies from the ALPINE survey are well resolved as the radial profiles of the stacks are more extended than the stacked synthesized beams (FWHM of $\sim1.1^{\prime\prime}$).
To compare the shape of radial profiles more directly, we created stacked [CII] images having the same beam size. We convolved the stacked images at $z\sim4.5$ and $z\sim5.5$ with Gaussians that have $\sigma^2 = \sigma^2_{z\sim7} - \sigma^2_{z\sim4/z\sim5}$, where $\sigma_{z\sim4/z\sim5/z\sim7}$ are the sigma of Gaussian fits to the respective beams of each stacked images. We note that the Gaussian convolutions resulted in  similar beams for all redshift bins.
We find that there is no noticeable difference in the radial profiles of all [CII] stacks for galaxies from $z\sim4.5$ to $z\sim7$ (Fig. \ref{fig:profile_compare}).

Additionally, to test a possible bias from the sample selection, we also stacked [CII] emission of $4<z<6$ galaxies  which have the same UV luminosity range as our $z\sim7$ galaxy sample.  
These UV luminosity matched sample have 39 galaxies at $4<z<6$, and have UV luminosity of $-23.0<M_{\rm UV}<-21.4$.
The stack of the UV luminosity matched $4<z<6$ sample shows $r_{\rm e}=2.17^{+0.16}_{-0.18}\,{\rm kpc}$, meaning that there is no clear change in the measured size of [CII] from $z\sim7$ (Tab. \ref{tab:stackresults}).
This suggests that the non-evolution we find is not strongly affected by sample selection in the different ALMA programs (Fig. \ref{fig:size-evolution}).

Overall, the measurements show that the average [CII] emission sizes have little or no evolution between $z\sim7$ and $z\sim4$ (Fig. \ref{fig:size-evolution}).
Although further high-resolution observations of a large sample is required to confirm  this trend (see \S\ref{sec:caveats}), the current measurements suggest that the non-evolution of [CII] emission sizes could be in contrast with the previously found evolution of the rest-UV continuum sizes, scaling as $\propto 1/(1+z)$ (e.g., \citealt{Shibuya2015}, however see also e.g., \citealt{Curtis-Lake2016} for possible non-evolution). If confirmed, the differential evolution may suggest that high-redshift galaxies might be morphologically dominated by gas, and star formation activity occupies a progressively smaller fraction of the volume in galaxies towards the highest redshifts.

\citet{Pizzati2020} showed that the extended [CII] emission can be explained by star formation driven outflow that have outflow velocity $\sim170\,\rm{km/s}$ and high mass-loading factor of $\eta=3.1$  defined as $\dot{M}=\eta\,{\rm SFR}$.
Also, the same hydrodynamical models tuned the velocity of winds with ALPINE measurements \citep{Ginolfi2020} predict a warm ISM of radius $r\sim2.5\,\rm{kpc}$ \citep[][]{Graziani2020}, in broad agreements with the results of the present work.

At $z\sim5$, the ALPINE survey detected the outflowing [CII] emission through a secondary broad line in the stacked [CII] spectrum, showing the velocity FWHM of $v_{\rm FWHM}\sim500-700\,\rm{km/s}$ \citep{Ginolfi2020}.
The finding of outflowing [CII] emission is further supported by an individual galaxy study of a star-forming galaxy at $z\sim5.54$, which shows an outflow and extended [CII] emission in a high-resolution ($\lesssim0.4^{\prime\prime}$) ALMA observation \citep{Herrera-Camus2021}.
Similar broad [CII] emission line is detected from the stacked [CII] 1D spectrum of the REBELS galaxies (Fudamoto et al. in prep).
The feature agrees with an outflow scenario of the extended [CII] emission in the previous observations \citep[e.g.,][]{Gallerani2018,Ginolfi2020} and the theoretical predictions \citep{Pizzati2020}.
A detailed analysis of the outflow properties in comparison with theoretical models will be presented in a future work.

\begin{table}
\caption{Stacked [CII] Sizes}
\centering
\begin{tabular}{cccc}
    \hline
    Redshift$^{\dagger}$ & \# of galaxies & $r_{\rm [CII]}$ [kpc] & $r_{\rm UV}$ [kpc]\\
    \hline\hline
    4.54 & 31 & $2.46^{+0.18}_{-0.18}$ & $1.06^{+0.38}_{-0.17}$$^{\ast}$\\
    5.57 & 21 & $2.23^{+0.25}_{-0.20}$ & $0.88^{+0.32}_{-0.07}$$^{\ast}$\\
    6.95 & 28 & $2.21^{+0.23}_{-0.19}$ & $0.85^{+0.16}_{-0.16}$ \\
    \hline\hline
    \multicolumn{4}{c}{$M_{\rm UV}$ matched $4<z<6$ galaxies{$^{\dagger\dagger}$}}\\
    \hline
    4.95 & 39 & $2.17^{+0.16}_{-0.18}$ & $1.04^{+0.27}_{-0.23}$$^{\ast}$\\
    \hline
    \multicolumn{4}{l}{$\dagger$ Median redshifts of each bin.}\\
    \multicolumn{4}{l}{$\dagger\dagger$ $-23.0<M_{\rm UV}<-21.4$ galaxies at $4<z<6$ (see \S\ref{sec:nonevolution})}\\
    \multicolumn{4}{l}{$\ast$ HST H-band size distribution in \citet{Fujimoto2020}}
\end{tabular}
\label{tab:stackresults}
\end{table}

\subsection{Potential Caveats}
\label{sec:caveats}
Our current analysis is based only on massive and highly star-forming galaxies that have [CII] emission lines and/or dust continuum detections.
\citet{Ginolfi2020} reported that [CII] emission sizes change as a function of star formation rate.
Similarly, rest-UV continuum observations show that the UV size strongly depends on the galaxies' UV luminosity \citep[e.g.,][]{Shibuya2015,Bowler2017,Bouwens2021b}. 
In particular, using high-resolution HST observations, \citet{Bowler2017} showed that the UV-brightest multi-component $z\sim7$  galaxies ($M_{\rm{UV}}\lesssim-21.5\,\rm{mag}$) have $r_{\rm e-UV}\sim1-3\,\rm{kpc}$, while single-component galaxies with $M_{\rm{UV}}\gtrsim-21.5\,\rm{mag}$ have $r_{\rm e-UV}\lesssim1\,\rm{kpc}$.
Our stacked rest-frame UV size ($r_{\rm{e-UV}}=0.83\pm0.07$) is consistent with the single-component galaxies in \citet{Bowler2017}.
However, the current spatial resolution of the J-band images (FWHM $\sim0.9^{\prime\prime}$) do not allow us to investigate further details of the rest-frame UV morphologies of our sample.
Our findings from the average size comparison between [CII] emission and rest-frame UV emission may not apply to individual galaxies, especially if multi-component galaxies exist in our sample.
Higher resolution images of the rest-frame UV emission will be required to study this in detail.

While we combined two ALMA large programs, the samples in each redshift bin only contain $\sim30$ galaxies, and all of them use relatively low resolution observations limiting our analysis to a small dynamic range and uncertain continuum size measurements.
Expanding the parameter space (e.g., higher angular resolutions and observations of lower mass galaxies) are required to confirm the gas size evolution of high-redshift galaxies.
Especially, higher angular resolution observation ($<1^{\prime\prime}$) will be crucial to provide more complete morpho-kinematic classifications, and  to avoid any possible uncertainties to size measurements by merging galaxies that we cannot find by $\sim1^{\prime\prime}$ resolution (e.g., late-stage mergers).

\section{Conclusions}
In this paper, we presented a study of the average [CII] $158\,\rm{\mu m}$ line emission, dust continuum, and rest-frame UV sizes of star-forming galaxies at $z\sim7$ based on the on-going ALMA large program REBELS.
We also estimate the average size of [CII] emission lines of $4<z<6$ star-forming galaxies using same stacking method to study the [CII] size evolution as a function of redshift. We summarize our findings:
\vspace{5pt}

\noindent(i) At $z\sim7$, the [CII] $158\,\rm{\mu m}$ emission line is spatially more extended than the dust continuum and the rest-frame UV emission.
We found the effective radius to be $r_{\rm{e}} = 2.21^{+0.23}_{-0.19}\,\rm{kpc}$, $1.14\pm0.27\,\rm{kpc}$, and $0.83\pm0.07\,\rm{kpc}$ for the [CII], dust continuum, and rest-frame UV emission, respectively.
The $\gtrsim2\times$ more extended [CII] emission tracing the gas is consistent with previous finding at $z\lesssim6$.
\vspace{5pt}

\noindent(ii) Comparing with the stacked [CII] emission sizes of galaxies from $z\sim4$ to $z\sim7$, we found little or no evolution of [CII] sizes in star-forming galaxies between $z\sim4$ and $z\sim7$.
If confirmed with further observations, the constant [CII] size could be in contrast with the previously found UV size evolution, and suggests that the [CII] emitting gas dominates the morphologies of high-redshift star-forming galaxies while star formation might occupy a progressively smaller fraction of size in galaxies towards high redshifts.
\vspace{5pt}

Further confirming this study would require larger samples of galaxies observed with ALMA, in particular expanding the redshift range observed to constrain the evolution of the gas sizes.
At the same time, higher resolution observations are essential to measure sizes more accurately, and to study detailed structures and morphologies of high-redshift star forming galaxies.
 
\vspace{5pt}
\begin{acknowledgments}

\noindent This paper makes use of the following ALMA data: 
\path{ADS/JAO.ALMA#2019.1.01634.L},
\path{ADS/JAO.ALMA#2018.1.00085.S},
\path{ADS/JAO.ALMA#2017.1.00428.L},
\path{ADS/JAO.ALMA#2015.1.01111.S}.
ALMA is a partnership of ESO (representing its member states), NSF(USA) and NINS (Japan), together with NRC (Canada), MOST and ASIAA (Taiwan), and KASI (Republic of Korea), in cooperation with the Republic of Chile. The Joint ALMA Observatory is operated by ESO, AUI/NRAO and NAOJ.
YF, YS, and AKI acknowledge support from NAOJ ALMA Scientific Research Grant number 2020-16B.
PAO and LB acknowledge support from the Swiss National Science Foundation through the SNSF Professorship grant 190079 ‘Galaxy Buildup at Cosmic Dawn’.
SS acknowledges support from the Nederlandse Onderzoekschool voor Astronomie (NOVA).
RS acknowledge support from STFC Ernest Rutherford Fellowshipsand  [grant number ST/S004831/1].
RB acknowledges support from an STFC Ernest Rutherford Fellowship [grant number ST/T003596/1].
RE acknowledges funding from JWST/NIRCam contract to the University of Arizona, NAS5-02015.
HI and HSBA acknowledge support from the NAOJ ALMA Scientific Research Grant Code 2021-19A. HI acknowledges support from the JSPS KAKENHI Grant Number JP19K23462.
JH gratefully acknowledges support of the VIDI research program with project number 639.042.611, which is (partly) financed by the Netherlands Organisation for Scientific Research (NWO). MA acknowledges support from FONDECYT grant 1211951, “CONICYT + PCI + INSTITUTO MAX PLANCK DE ASTRONOMIA MPG190030” and “CONICYT+PCI+REDES 190194” and ANID BASAL project FB210003.
P. Dayal acknowledges support from the European Research Council's starting grant ERC StG-717001 (``DELPHI"), from the NWO grant 016.VIDI.189.162 (``ODIN") and the European Commission's and University of Groningen's CO-FUND Rosalind Franklin program.
LG and RS acknowledge support from the Amaldi Research Center funded by the MIUR program “Dipartimento di Eccellenza” (CUP:B81I18001170001).
 AF acknowledges support from the ERC Advanced Grant INTERSTELLAR H2020/740120.
Partial support from the Carl Friedrich von Siemens-Forschungspreis der Alexander von Humboldt-Stiftung Research Award is kindly acknowledged (AF).
IDL acknowledges support from ERC starting grant 851622 DustOrigin.
JW acknowledges support from the ERC Advanced Grant 695671, “QUENCH”, and from the Fondation MERAC.
GCJ acknowledges funding from the ``FirstGalaxies'' Advanced Grant from the European Research Council (ERC) under the European Union’s Horizon 2020 research and innovation programme (Grant agreement No. 789056).
The Cosmic Dawn Center (DAWN) is funded by the Danish National Research Foundation under grant No. 140.
\end{acknowledgments}

\bibliography{base}{}

\begin{thebibliography}{}
\expandafter\ifx\csname natexlab\endcsname\relax\def\natexlab#1{#1}\fi
\providecommand{\url}[1]{\href{#1}{#1}}
\providecommand{\dodoi}[1]{doi:~\href{http://doi.org/#1}{\nolinkurl{#1}}}
\providecommand{\doeprint}[1]{\href{http://ascl.net/#1}{\nolinkurl{http://ascl.net/#1}}}
\providecommand{\doarXiv}[1]{\href{https://arxiv.org/abs/#1}{\nolinkurl{https://arxiv.org/abs/#1}}}

\bibitem[{{B{\'e}thermin} {et~al.}(2020){B{\'e}thermin}, {Fudamoto}, {Ginolfi},
  {Loiacono}, {Khusanova}, {Capak}, {Cassata}, {Faisst}, {Le F{\`e}vre},
  {Schaerer}, {Silverman}, {Yan}, {Amorin}, {Bardelli}, {Boquien}, {Cimatti},
  {Davidzon}, {Dessauges-Zavadsky}, {Fujimoto}, {Gruppioni}, {Hathi}, {Ibar},
  {Jones}, {Koekemoer}, {Lagache}, {Lemaux}, {Moreau}, {Oesch}, {Pozzi},
  {Riechers}, {Talia}, {Toft}, {Vallini}, {Vergani}, {Zamorani}, \&
  {Zucca}}]{Bethermin2020}
{B{\'e}thermin}, M., {Fudamoto}, Y., {Ginolfi}, M., {et~al.} 2020, \aap, 643,
  A2, \dodoi{10.1051/0004-6361/202037649}

\bibitem[{{Bigiel} \& {Blitz}(2012)}]{Bigiel2012}
{Bigiel}, F., \& {Blitz}, L. 2012, \apj, 756, 183,
  \dodoi{10.1088/0004-637X/756/2/183}

\bibitem[{{Bouwens} {et~al.}(2021{\natexlab{a}}){Bouwens}, {Illingworth}, {van
  Dokkum}, {Ribeiro}, {Oesch}, \& {Stefanon}}]{Bouwens2021b}
{Bouwens}, R.~J., {Illingworth}, G.~D., {van Dokkum}, P.~G., {et~al.}
  2021{\natexlab{a}}, \aj, 162, 255, \dodoi{10.3847/1538-3881/abfda6}

\bibitem[{{Bouwens} {et~al.}(2021{\natexlab{b}}){Bouwens}, {Smit}, {Schouws},
  {Stefanon}, {Bowler}, {Endsley}, {Gonzalez}, {Inami}, {Stark}, {Oesch},
  {Hodge}, {Aravena}, {da Cunha}, {Dayal}, {de Looze}, {Ferrara}, {Fudamoto},
  {Graziani}, {Li}, {Nanayakkara}, {Pallotini}, {Schneider}, {Sommovigo},
  {Topping}, {van der Werf}, {Barrufet}, {Hygate}, {Labbe}, {Riechers}, \&
  {Witstok}}]{Bouwens2021}
{Bouwens}, R.~J., {Smit}, R., {Schouws}, S., {et~al.} 2021{\natexlab{b}}, arXiv
  e-prints, arXiv:2106.13719.
\newblock \doarXiv{2106.13719}

\bibitem[{{Bowler} {et~al.}(2022){Bowler}, {Cullen}, {McLure}, {Dunlop}, \&
  {Avison}}]{Bowler2022}
{Bowler}, R.~A.~A., {Cullen}, F., {McLure}, R.~J., {Dunlop}, J.~S., \&
  {Avison}, A. 2022, \mnras, 510, 5088, \dodoi{10.1093/mnras/stab3744}

\bibitem[{{Bowler} {et~al.}(2017){Bowler}, {Dunlop}, {McLure}, \&
  {McLeod}}]{Bowler2017}
{Bowler}, R.~A.~A., {Dunlop}, J.~S., {McLure}, R.~J., \& {McLeod}, D.~J. 2017,
  \mnras, 466, 3612, \dodoi{10.1093/mnras/stw3296}

\bibitem[{{Chabrier}(2003)}]{Chabrier2003}
{Chabrier}, G. 2003, \pasp, 115, 763, \dodoi{10.1086/376392}

\bibitem[{{Curtis-Lake} {et~al.}(2016){Curtis-Lake}, {McLure}, {Dunlop},
  {Rogers}, {Targett}, {Dekel}, {Ellis}, {Faber}, {Ferguson}, {Grogin},
  {Kocevski}, {Koekemoer}, {Lai}, {M{\'a}rmol-Queralt{\'o}}, \&
  {Robertson}}]{Curtis-Lake2016}
{Curtis-Lake}, E., {McLure}, R.~J., {Dunlop}, J.~S., {et~al.} 2016, \mnras,
  457, 440, \dodoi{10.1093/mnras/stv3017}

\bibitem[{{Dayal} {et~al.}(2014){Dayal}, {Ferrara}, {Dunlop}, \&
  {Pacucci}}]{Dayal2014}
{Dayal}, P., {Ferrara}, A., {Dunlop}, J.~S., \& {Pacucci}, F. 2014, \mnras,
  445, 2545, \dodoi{10.1093/mnras/stu1848}

\bibitem[{{Decarli} {et~al.}(2020){Decarli}, {Aravena}, {Boogaard}, {Carilli},
  {Gonz{\'a}lez-L{\'o}pez}, {Walter}, {Cortes}, {Cox}, {da Cunha}, {Daddi},
  {D{\'\i}az-Santos}, {Hodge}, {Inami}, {Neeleman}, {Novak}, {Oesch},
  {Popping}, {Riechers}, {Smail}, {Uzgil}, {van der Werf}, {Wagg}, \&
  {Weiss}}]{Decarli2020}
{Decarli}, R., {Aravena}, M., {Boogaard}, L., {et~al.} 2020, \apj, 902, 110,
  \dodoi{10.3847/1538-4357/abaa3b}

\bibitem[{{Dessauges-Zavadsky} {et~al.}(2020){Dessauges-Zavadsky}, {Ginolfi},
  {Pozzi}, {B{\'e}thermin}, {Le F{\`e}vre}, {Fujimoto}, {Silverman}, {Jones},
  {Vallini}, {Schaerer}, {Faisst}, {Khusanova}, {Fudamoto}, {Cassata},
  {Loiacono}, {Capak}, {Yan}, {Amorin}, {Bardelli}, {Boquien}, {Cimatti},
  {Gruppioni}, {Hathi}, {Ibar}, {Koekemoer}, {Lemaux}, {Narayanan}, {Oesch},
  {Rodighiero}, {Romano}, {Talia}, {Toft}, {Vergani}, {Zamorani}, \&
  {Zucca}}]{Dessauges2020}
{Dessauges-Zavadsky}, M., {Ginolfi}, M., {Pozzi}, F., {et~al.} 2020, \aap, 643,
  A5, \dodoi{10.1051/0004-6361/202038231}

\bibitem[{{Faisst} {et~al.}(2020){Faisst}, {Schaerer}, {Lemaux}, {Oesch},
  {Fudamoto}, {Cassata}, {B{\'e}thermin}, {Capak}, {Le F{\`e}vre}, {Silverman},
  {Yan}, {Ginolfi}, {Koekemoer}, {Morselli}, {Amor{\'\i}n}, {Bardelli},
  {Boquien}, {Brammer}, {Cimatti}, {Dessauges-Zavadsky}, {Fujimoto},
  {Gruppioni}, {Hathi}, {Hemmati}, {Ibar}, {Jones}, {Khusanova}, {Loiacono},
  {Pozzi}, {Talia}, {Tasca}, {Riechers}, {Rodighiero}, {Romano}, {Scoville},
  {Toft}, {Vallini}, {Vergani}, {Zamorani}, \& {Zucca}}]{Faisst2020}
{Faisst}, A.~L., {Schaerer}, D., {Lemaux}, B.~C., {et~al.} 2020, \apjs, 247,
  61, \dodoi{10.3847/1538-4365/ab7ccd}

\bibitem[{{Fujimoto} {et~al.}(2019){Fujimoto}, {Ouchi}, {Ferrara},
  {Pallottini}, {Ivison}, {Behrens}, {Gallerani}, {Arata}, {Yajima}, \&
  {Nagamine}}]{Fujimoto2019}
{Fujimoto}, S., {Ouchi}, M., {Ferrara}, A., {et~al.} 2019, \apj, 887, 107,
  \dodoi{10.3847/1538-4357/ab480f}

\bibitem[{{Fujimoto} {et~al.}(2020){Fujimoto}, {Silverman}, {Bethermin},
  {Ginolfi}, {Jones}, {Le F{\`e}vre}, {Dessauges-Zavadsky}, {Rujopakarn},
  {Faisst}, {Fudamoto}, {Cassata}, {Morselli}, {Maiolino}, {Schaerer}, {Capak},
  {Yan}, {Vallini}, {Toft}, {Loiacono}, {Zamorani}, {Talia}, {Narayanan},
  {Hathi}, {Lemaux}, {Boquien}, {Amorin}, {Ibar}, {Koekemoer},
  {M{\'e}ndez-Hern{\'a}ndez}, {Bardelli}, {Vergani}, {Zucca}, {Romano}, \&
  {Cimatti}}]{Fujimoto2020}
{Fujimoto}, S., {Silverman}, J.~D., {Bethermin}, M., {et~al.} 2020, \apj, 900,
  1, \dodoi{10.3847/1538-4357/ab94b3}

\bibitem[{{Gallerani} {et~al.}(2018){Gallerani}, {Pallottini}, {Feruglio},
  {Ferrara}, {Maiolino}, {Vallini}, {Riechers}, \& {Pavesi}}]{Gallerani2018}
{Gallerani}, S., {Pallottini}, A., {Feruglio}, C., {et~al.} 2018, \mnras, 473,
  1909, \dodoi{10.1093/mnras/stx2458}

\bibitem[{{Ginolfi} {et~al.}(2020){Ginolfi}, {Jones}, {B{\'e}thermin},
  {Fudamoto}, {Loiacono}, {Fujimoto}, {Le F{\'e}vre}, {Faisst}, {Schaerer},
  {Cassata}, {Silverman}, {Yan}, {Capak}, {Bardelli}, {Boquien}, {Carraro},
  {Dessauges-Zavadsky}, {Giavalisco}, {Gruppioni}, {Ibar}, {Khusanova},
  {Lemaux}, {Maiolino}, {Narayanan}, {Oesch}, {Pozzi}, {Rodighiero}, {Talia},
  {Toft}, {Vallini}, {Vergani}, \& {Zamorani}}]{Ginolfi2020}
{Ginolfi}, M., {Jones}, G.~C., {B{\'e}thermin}, M., {et~al.} 2020, \aap, 633,
  A90, \dodoi{10.1051/0004-6361/201936872}

\bibitem[{{Graziani} {et~al.}(2020){Graziani}, {Schneider}, {Ginolfi}, {Hunt},
  {Maio}, {Glatzle}, \& {Ciardi}}]{Graziani2020}
{Graziani}, L., {Schneider}, R., {Ginolfi}, M., {et~al.} 2020, \mnras, 494,
  1071, \dodoi{10.1093/mnras/staa796}

\bibitem[{{Herrera-Camus} {et~al.}(2021){Herrera-Camus}, {F{\"o}rster
  Schreiber}, {Genzel}, {Tacconi}, {Bolatto}, {Davies}, {Fisher}, {Lutz},
  {Naab}, {Shimizu}, {Tadaki}, \& {{\"U}bler}}]{Herrera-Camus2021}
{Herrera-Camus}, R., {F{\"o}rster Schreiber}, N., {Genzel}, R., {et~al.} 2021,
  \aap, 649, A31, \dodoi{10.1051/0004-6361/202039704}

\bibitem[{{Hodge} \& {da Cunha}(2020)}]{Hodge2020}
{Hodge}, J.~A., \& {da Cunha}, E. 2020, Royal Society Open Science, 7, 200556,
  \dodoi{10.1098/rsos.200556}

\bibitem[{{Inami} {et~al.}(2022){Inami}, {Algera}, {Schouws}, {Sommovigo},
  {Bouwens}, {Smit}, {Stefanon}, {Bowler}, {Endsley}, {Ferrara}, {Oesch},
  {Stark}, {Aravena}, {Barrufet}, {da Cunha}, {Dayal}, {De Looze}, {Fudamoto},
  {Gonzalez}, {Graziani}, {Hodge}, {Hygate}, {Nanayakkara}, {Pallottini},
  {Riechers}, {Schneider}, {Topping}, \& {van der Werf}}]{Inami2022}
{Inami}, H., {Algera}, H. S.~B., {Schouws}, S., {et~al.} 2022, arXiv e-prints,
  arXiv:2203.15136.
\newblock \doarXiv{2203.15136}

\bibitem[{{Jarvis} {et~al.}(2013){Jarvis}, {Bonfield}, {Bruce}, {Geach},
  {McAlpine}, {McLure}, {Gonz{\'a}lez-Solares}, {Irwin}, {Lewis}, {Yoldas},
  {Andreon}, {Cross}, {Emerson}, {Dalton}, {Dunlop}, {Hodgkin}, {Le},
  {Karouzos}, {Meisenheimer}, {Oliver}, {Rawlings}, {Simpson}, {Smail},
  {Smith}, {Sullivan}, {Sutherland}, {White}, \& {Zwart}}]{Jarvis2013}
{Jarvis}, M.~J., {Bonfield}, D.~G., {Bruce}, V.~A., {et~al.} 2013, \mnras, 428,
  1281, \dodoi{10.1093/mnras/sts118}

\bibitem[{{Le F{\`e}vre} {et~al.}(2020){Le F{\`e}vre}, {B{\'e}thermin},
  {Faisst}, {Jones}, {Capak}, {Cassata}, {Silverman}, {Schaerer}, {Yan},
  {Amorin}, {Bardelli}, {Boquien}, {Cimatti}, {Dessauges-Zavadsky},
  {Giavalisco}, {Hathi}, {Fudamoto}, {Fujimoto}, {Ginolfi}, {Gruppioni},
  {Hemmati}, {Ibar}, {Koekemoer}, {Khusanova}, {Lagache}, {Lemaux}, {Loiacono},
  {Maiolino}, {Mancini}, {Narayanan}, {Morselli}, {M{\'e}ndez-Hern{\`a}ndez},
  {Oesch}, {Pozzi}, {Romano}, {Riechers}, {Scoville}, {Talia}, {Tasca},
  {Thomas}, {Toft}, {Vallini}, {Vergani}, {Walter}, {Zamorani}, \&
  {Zucca}}]{Lefevre2020}
{Le F{\`e}vre}, O., {B{\'e}thermin}, M., {Faisst}, A., {et~al.} 2020, \aap,
  643, A1, \dodoi{10.1051/0004-6361/201936965}

\bibitem[{{Liu} {et~al.}(2019){Liu}, {Schinnerer}, {Groves}, {Magnelli},
  {Lang}, {Leslie}, {Jim{\'e}nez-Andrade}, {Riechers}, {Popping}, {Magdis},
  {Daddi}, {Sargent}, {Gao}, {Fudamoto}, {Oesch}, \& {Bertoldi}}]{Liu2019}
{Liu}, D., {Schinnerer}, E., {Groves}, B., {et~al.} 2019, \apj, 887, 235,
  \dodoi{10.3847/1538-4357/ab578d}

\bibitem[{{Madau} \& {Dickinson}(2014)}]{Madau2014}
{Madau}, P., \& {Dickinson}, M. 2014, \araa, 52, 415,
  \dodoi{10.1146/annurev-astro-081811-125615}

\bibitem[{{Mart{\'\i}-Vidal} {et~al.}(2014){Mart{\'\i}-Vidal}, {Vlemmings},
  {Muller}, \& {Casey}}]{Marti-Vidal2014}
{Mart{\'\i}-Vidal}, I., {Vlemmings}, W.~H.~T., {Muller}, S., \& {Casey}, S.
  2014, \aap, 563, A136, \dodoi{10.1051/0004-6361/201322633}

\bibitem[{{McCracken} {et~al.}(2012){McCracken}, {Milvang-Jensen}, {Dunlop},
  {Franx}, {Fynbo}, {Le F{\`e}vre}, {Holt}, {Caputi}, {Goranova}, {Buitrago},
  {Emerson}, {Freudling}, {Hudelot}, {L{\'o}pez-Sanjuan}, {Magnard}, {Mellier},
  {M{\o}ller}, {Nilsson}, {Sutherland}, {Tasca}, \& {Zabl}}]{McCracken2012}
{McCracken}, H.~J., {Milvang-Jensen}, B., {Dunlop}, J., {et~al.} 2012, \aap,
  544, A156, \dodoi{10.1051/0004-6361/201219507}

\bibitem[{{Peng} {et~al.}(2010){Peng}, {Ho}, {Impey}, \& {Rix}}]{Pang2010}
{Peng}, C.~Y., {Ho}, L.~C., {Impey}, C.~D., \& {Rix}, H.-W. 2010, \aj, 139,
  2097, \dodoi{10.1088/0004-6256/139/6/2097}

\bibitem[{{Pizzati} {et~al.}(2020){Pizzati}, {Ferrara}, {Pallottini},
  {Gallerani}, {Vallini}, {Decataldo}, \& {Fujimoto}}]{Pizzati2020}
{Pizzati}, E., {Ferrara}, A., {Pallottini}, A., {et~al.} 2020, \mnras, 495,
  160, \dodoi{10.1093/mnras/staa1163}

\bibitem[{{Romano} {et~al.}(2021){Romano}, {Cassata}, {Morselli}, {Jones},
  {Ginolfi}, {Zanella}, {B{\'e}thermin}, {Capak}, {Faisst}, {Le F{\`e}vre},
  {Schaerer}, {Silverman}, {Yan}, {Bardelli}, {Boquien}, {Cimatti},
  {Dessauges-Zavadsky}, {Enia}, {Fujimoto}, {Gruppioni}, {Hathi}, {Ibar},
  {Koekemoer}, {Lemaux}, {Rodighiero}, {Vergani}, {Zamorani}, \&
  {Zucca}}]{Romano2021}
{Romano}, M., {Cassata}, P., {Morselli}, L., {et~al.} 2021, \aap, 653, A111,
  \dodoi{10.1051/0004-6361/202141306}

\bibitem[{{Schouws} {et~al.}(2021){Schouws}, {Stefanon}, {Bouwens}, {Smit},
  {Hodge}, {Labb{\'e}}, {Algera}, {Boogaard}, {Carniani}, {Fudamoto},
  {Holwerda}, {Illingworth}, {Maiolino}, {Maseda}, {Oesch}, \& {van der
  Werf}}]{Schouws2021}
{Schouws}, S., {Stefanon}, M., {Bouwens}, R.~J., {et~al.} 2021, arXiv e-prints,
  arXiv:2105.12133.
\newblock \doarXiv{2105.12133}

\bibitem[{{Schouws} {et~al.}(2022){Schouws}, {Bouwens}, {Smit}, {Hodge},
  {Stefanon}, {Witstok}, {Hilhorst}, {Labbe}, {Algera}, {Boogaard}, {Maseda},
  {Oesch}, {R{\"o}ttgering}, \& {van der Werf}}]{Schouws2022}
{Schouws}, S., {Bouwens}, R., {Smit}, R., {et~al.} 2022, arXiv e-prints,
  arXiv:2202.04080.
\newblock \doarXiv{2202.04080}

\bibitem[{{Shibuya} {et~al.}(2015){Shibuya}, {Ouchi}, \&
  {Harikane}}]{Shibuya2015}
{Shibuya}, T., {Ouchi}, M., \& {Harikane}, Y. 2015, \apjs, 219, 15,
  \dodoi{10.1088/0067-0049/219/2/15}

\bibitem[{{Smit} {et~al.}(2018){Smit}, {Bouwens}, {Carniani}, {Oesch},
  {Labb{\'e}}, {Illingworth}, {van der Werf}, {Bradley}, {Gonzalez}, {Hodge},
  {Holwerda}, {Maiolino}, \& {Zheng}}]{Smit2018}
{Smit}, R., {Bouwens}, R.~J., {Carniani}, S., {et~al.} 2018, \nat, 553, 178,
  \dodoi{10.1038/nature24631}

\bibitem[{{Sommovigo} {et~al.}(2022){Sommovigo}, {Ferrara}, {Pallottini},
  {Dayal}, {Bouwens}, {Smit}, {da Cunha}, {De Looze}, {Bowler}, {Hodge},
  {Inami}, {Oesch}, {Endsley}, {Gonzalez}, {Schouws}, {Stark}, {Stefanon},
  {Aravena}, {Graziani}, {Riechers}, {Schneider}, {van der Werf}, {Algera},
  {Barrufet}, {Fudamoto}, {Hygate}, {Labb{\'e}}, {Li}, {Nanayakkara}, \&
  {Topping}}]{Sommovigo2022}
{Sommovigo}, L., {Ferrara}, A., {Pallottini}, A., {et~al.} 2022, arXiv
  e-prints, arXiv:2202.01227.
\newblock \doarXiv{2202.01227}

\bibitem[{{Tacconi} {et~al.}(2020){Tacconi}, {Genzel}, \&
  {Sternberg}}]{Tacconi2020}
{Tacconi}, L.~J., {Genzel}, R., \& {Sternberg}, A. 2020, \araa, 58, 157,
  \dodoi{10.1146/annurev-astro-082812-141034}

\bibitem[{{Wang} {et~al.}(2022){Wang}, {Magnelli}, {Schinnerer}, {Liu}, {Aziz
  Modak}, {Faustino Jim{\'e}nez-Andrade}, {Karoumpis}, {Kokorev}, \&
  {Bertoldi}}]{Wang2022}
{Wang}, T.-M., {Magnelli}, B., {Schinnerer}, E., {et~al.} 2022, arXiv e-prints,
  arXiv:2201.12070.
\newblock \doarXiv{2201.12070}

\end{thebibliography}
\bibliographystyle{aasjournal}



\end{document}